\documentclass[11pt,preprint,aps,superscriptaddress,nofootinbib]{revtex4}

\usepackage{amsmath, amssymb,graphics, graphicx, bm}

\begin{document}

\preprint{UCB-PTH-08/67}

\title{A Star Product for Differential Forms on Symplectic Manifolds 
\footnote{This work was supported in part by the Director, Office of Science,
Office of High Energy and Nuclear Physics, Division of High Energy
Physics of the U.S. Department of Energy under Contract
DE-AC03-76SF00098, in part by the National Science Foundation under
grant PHY-0457315.} }

\author{Anthony Tagliaferro}
\email{atag@berkeley.edu}
\affiliation{Department of Physics, University of California, Berkeley}
\affiliation{Theoretical Physics Group, LBNL, Berkeley, CA 94702, USA}


\date{\today}

\newcommand{\R}{\ensuremath{\mathbb{R}}}
\newcommand{\C}{\ensuremath{\mathbb{C}}}
\renewcommand{\H}{\ensuremath{\mathbb{H}}}
\newcommand{\Z}{\ensuremath{\mathbb{Z}}}
\newcommand{\D}{\ensuremath{\mathcal{D}}}
\newcommand{\N}{\ensuremath{\mathcal{N}}}
\newcommand{\g}{{\ensuremath{\gamma}}}
\newcommand{\h}{\ensuremath{\mathfrak{h}}}
\newcommand{\z}{{\ensuremath{\bar{z}}}}
\newcommand{\fig}[2]{ \begin{figure}[h] \begin{center} \includegraphics[width=0.75\textwidth]{#1} \caption{#2} \label{#1} \end{center} \end{figure}}
\newcommand{\smallfig}[2]{ \begin{figure}[h] \begin{center} \includegraphics[width=0.5\textwidth]{#1} \caption{#2} \label{#1} \end{center} \end{figure}}
\newcommand{\ket}[1]{\ensuremath{|#1 \rangle}}
\newcommand{\bra}[1]{\ensuremath{\langle #1|}}
\newcommand{\braket}[2]{\ensuremath{\left\langle #1|#2 \right\rangle}}
\newcommand{\corr}[1]{\ensuremath{\left\langle #1 \right\rangle}}
\newcommand{\be}{\begin{equation}}
\newcommand{\ee}{\end{equation}}
\renewcommand{\align}[1]{\begin{align*} #1 \end{align*}}
\renewcommand{\d}{\ensuremath{\partial}}
\renewcommand{\d}{\ensuremath{\partial}}
\renewcommand{\(}{\ensuremath{\left( }}
\renewcommand{\)}{\ensuremath{\right) }}
\renewcommand{\[}{\ensuremath{\left[}}
\renewcommand{\]}{\ensuremath{\right] }}
\renewcommand{\t}[1]{\widetilde{#1}}
\renewcommand{\a}{\ensuremath{\alpha}}
\renewcommand{\b}{\ensuremath{\beta}}
\newcommand{\q}{\ensuremath{\theta}}
\newcommand{\e}{\ensuremath{\epsilon}}
\newcommand{\cH}{\ensuremath{\mathcal{H}}}
\newcommand{\cL}{\ensuremath{\mathcal{L}}}
\newcommand{\cO}{\ensuremath{\mathcal{O}}}
\newcommand{\cE}{\ensuremath{\mathcal{E}}}
\newcommand{\cG}{\ensuremath{\mathcal{G}}}
\newcommand{\cA}{\ensuremath{\mathcal{A}}}

\setlength{\parindent}{0 pt}
\setlength{\parskip}{5 pt}

\begin{abstract}
We present a star product between differential forms to second order
in the deformation parameter $\hbar$. The star product obtained is
consistent with a graded differential
Poisson algebra structure on a symplectic manifold.  The form of the
graded differential Poisson algebra requires the introduction of a
connection with torsion on the manifold, and places various
constraints upon it. The star product is given to second order in
$\hbar^2$.
\end{abstract}

\maketitle


\section{Introduction}

Since the early days of quantum mechanics, physicists have used star
products to build noncommutative generalizations of commuting theories
(see \cite{ds} for a historical treatment and the references therein).
From describing theories on noncommutative spaces to providing
alternate methods of quantization \cite{f}, star products and, more
generally, the procedure of deformation quantization have given
physicists a way to generalize commutative theories into
noncommutative theories.

Deformation quantization begins with a manifold $M$ and the ring of
smooth functions $C^\infty(M)$.  The ring of commutative smooth
functions $C^\infty(M)$ is deformed to a noncommutative ring of smooth
functions, $C^\infty(M,*)[[\hbar]]$, where the deformation is
parametrized by $\hbar$. We denote the noncommutative product of
$f,g\in C^\infty(M,*)[[\hbar]]$ by the star product:
$$f*g = fg +\hbar \{f,g\}+\hbar^2C_2(f,g)+\cO(\hbar^3)$$ where
$\{f,g\}=\pi^{ij}\d_if\d_jg$ is a Poisson bracket on $M$.  Smooth
functions $f,g\in C^\infty(M)$ are elements of the noncommutative
algebra, and in fact, as a set $C^\infty(M)\subset
C^\infty(M,*)[[\hbar]]$, but with a new product between elements.
Although $C^\infty(M,*)[[\hbar]]$ is not commutative, it is required
to be associative.  Associativity places strong conditions on the form
of the star product. The star product of functions is well known from
the work of Kontsevich \cite{k}, Cattaneo and Felder \cite{cf}, and
many others, including \cite{kv}.  We generalize the star product to
act between differential forms, to $\mathcal{O}(\hbar^2)$.

Gauge theories involve the differential form $A_\mu dx^\mu$, so to
study gauge theories on noncommutative spaces, the star product must
be extended to include differential forms.  With this motivation, we
deform the graded exterior algebra of differential forms,
$\Omega^*(M)$, to a noncommutative graded exterior algebra
$\Omega^*(M,*)[[\hbar]]$.  Thinking of functions as 0-forms, we need
to extend the star product to include forms of arbitrary degree:
$$ \a * \b = \a\b + \hbar \{\a,\b\}+\hbar^2C_2(\a,\b)+\cO(\hbar^3)$$
where $\{\a,\b\}$ is the generalization of the Poisson bracket to
differential forms, and $\a\b$ denotes the wedge
product\footnote{Unless otherwise specified, the product between two
differential forms is the wedge product: $\a\b=\a\wedge\b$}.  The case
of constant Poisson bivector, $\pi^{ij}$ is well understood, but we
are interested in the case that $\pi^{ij}$ is not constant.  Modulo
some conditions to be discussed later, it is possible to define a
consistent star product on the space of differential forms.
 
In this note we give the explicit form of the star product for
differential forms for nonconstant $\pi^{ij}$, to
$\mathcal{O}(\hbar^2)$.  First we must generalize the Poisson bracket
to differential forms.  To do so, we write down the properties a
graded differential Poisson algebra should satisfy, following
\cite{ch} and \cite{hm}.  After some motivation, we define an explicit
graded Poisson bracket between arbitrary forms, and check that it
satisfies the properties required of a graded differential Poisson
bracket.  We show that these properties place several conditions on
our manifold (these conditions were also found in \cite{ch} and
\cite{hm}).  Lastly, we propose an $\mathcal{O}(\hbar^2)$ product
between differential forms and show that it satisfies the properties
of a star product to that order.  In particular, we show explicitly
that our proposed product is associative, a crucial property of the
star product.

\section{Properties of a Differential Poisson Algebra}

The Poisson bracket between two functions, $\{f,g\}$, is ubiquitous in
the theory of classical mechanics, and its properties are widely
known (see \cite{v} among many others).  The basic properties are:
\begin{enumerate}
\item Skew-symmetry: $\{f,g\} = -\{g, f\}$ 
\item Jacobi identity: $\{f, \{g, h\}\}+\{g, \{h, f\}\} + \{h, \{g, f\}\}=0$
\item Leibniz rule: $\{f, g h\}= \{f, g\}h + g\{f,h\}.$
\end{enumerate}
Define the Poisson bivector $\pi^{ij}$ by:
\begin{equation}\{f,g\} = \pi^{ij}\d_if\d_jg.\end{equation}
Because the Poisson bracket obeys the Jacobi identity, $\pi^{ij}$ must
satisfy the following condition:
\begin{equation}\pi^{im}\d_m\pi^{jk}+\pi^{jm}\d_m\pi^{ki}+\pi^{km}\d_m\pi^{ij}=0.\label{jac}
\end{equation} 
Since functions are 0-forms, any Poisson bracket on the space of
differential forms, $\Omega^*(M)$, must reduce to this simple Poisson
bracket when restricted to 0-forms.

When a Poisson bracket is defined on a manifold, the manifold is called
a Poisson manifold. If the Poisson bivector is invertible, 
its inverse, $\omega_{ij}$, is called the symplectic 2-form and the manifold a symplectic 
manifold. We use the convention:
$$\pi^{ij}\omega_{jk}=\delta^i_{\phantom{j}k}$$ Equation (\ref{jac})  
is equivalent to the condition $d\omega = 0$, where
$\omega=\frac12\omega_{ij}dx^idx^j$. For simplicity, we take the
manifold $M$ to be symplectic, but the results may generalize to the
more general Poisson manifold case.

Following \cite{ch}, we generalize the Poisson bracket to include
differential forms.  Denote by $|\a|$ the degree of the form $\a$.
For all forms $\a, \b,$ and $\g$, a graded differential Poisson
algebra on a Poisson manifold must satisfy:
\begin{enumerate}
\item  Bracket degree:  \label{propbd}
\be |\{\a,\b\}|=|\a|+|\b| \label{BD}\ee
\item Graded symmetry: 
\be\{\a,\b\}=(-1)^{|\a||\b|+1}\{\b,\a\}\ee
\item Graded product rule: 
\be\{\a,\b\g\}=\{\a,\b\}\g+(-1)^{|\a||\b|}\b\{\a,\g\} \ee
\item Leibniz rule for the exterior derivative, $d$: 
\be d\{\a,\b\}=\{d\a,\b\}+(-1)^{|\a|}\{\a,d\b\} \ee
\item Graded Jacobi identity: \label{propgraded}
\be\{\a,\{\b,\g\}\} +(-1)^{|\a|(|\b|+|\g|)}\{\b,\{\g,\a\}\}+(-1)^{|\g|(|\a|+|\b|)}\{\g,\{\a,\b\}\}=0 \label{Graded} \ee
\end{enumerate}

These properties naturally combine the defining characteristics of
differential forms and the Poisson bracket. The Leibniz rule and the
Jacobi identity place strong constraints on how the Poisson bracket
acts on differential forms.  As we show in the next section,
Properties \ref{propbd}-\ref{propgraded} uniquely determine the form
of the Poisson bracket.

\section{The Poisson Bracket}

In this section, we motivate the form of the Poisson bracket between
two arbitrary differential forms on a symplectic manifold. We start by
using Properties \ref{propbd}-\ref{propgraded} to determine the Poisson bracket
between a function and an arbitrary differential form.  Next, we introduce the most general Poisson bracket between two
differential forms and discuss some of the conditions that arise. More
detailed calculations are included in Appendix \ref{Ver1}.

\subsection{The Poisson Bracket between a function and a form}

In this section, we show that for a symplectic manifold, there is
a unique Poisson bracket consistent with the
Poisson algebra.  In particular, we show that the object
$\{f,\cdot\}$ acts as a covariant derivative on the space of
differential forms:
$$\{ f, \a\} = \nabla_V\a$$ 
To be a connection, $\nabla_V\a$ must satisfy the following
properties for $f, g$ functions, $V$ a vector field, and a form $\a$:

\begin{center}\begin{tabular}{ll}
1. $\nabla_V(f) = V(f)$
& 4. $\nabla_{(V+W)}\a = \nabla_V\a +\nabla_W\a$ \\
2. $\nabla_V(f\a) = V(f)\a + f \nabla_V\a$ 
& 5. $\nabla_V(\a\wedge\b)=(\nabla_V\a)\wedge \b + \a \wedge(\nabla_V\b)$ \\
3. $\nabla_V(\a+\b)=\nabla_V\a + \nabla_V\b\qquad\qquad\quad$
& 6. $\nabla_{(gV)}\a = g\nabla_V\a.$\\
\end{tabular}\end{center}

The Poisson bracket $\{f,\cdot\}$ satisfies these properties with
$V^j=\pi^{ij}\d_if$.  Multiplying $V^j= \pi^{ij}\d_if= -\pi^{ji}\d_if
$ by $\omega_{jk}$, we have
$$ \d_kf = \omega_{jk}V^j, $$
which has as the solution
$$f= \int_{x_0}^x \omega_{jk} V^j dx^k.$$

Before checking that the bracket $\{f,\a\}$ is a covariant derivative
of the second argument, note that it is a differential operator of
order 1 in the first argument (e.g. $\{1,\a\}=0$ and
$\{fg,\a\}=\{f,\a\}g+f\{g,\a\}$). This implies:
$$\{f,\a\} = (\d_i f)\{x^i,\a\}.$$
A simple proof follows from Taylor expanding $f$ inside the Poisson bracket and noticing that only
the term linear in $x^i$ does not vanish:
$$\{f,\a\}\Big\vert_{x=x_0} =\lim_{x\to x_0}\bigg[ \{f(x_0),\a\} +
\(\d_if(x_0)\)\{x^i-x_0^i,\a\}+\(\frac12
\d_i\d_jf(x_0)\)\{(x^i-x_0^i)(x^j-x_0^j),\a\}+\dots\bigg].$$ From
this, we see that $\{f,\a\}=V^i\omega_{ij}\{x^j,\a\}$, which makes it
easy to prove that $\{f,\cdot\}$ satisfies all the properties of a
covariant derivative.  Note that our arguments depend on the existence
of $\omega_{ij}$. Although a general Poisson manifold, for which
$\pi^{ij}$ is not invertible, might have a different Poisson bracket,
we do not consider this case.

Inserting the vector field $V$, the Poisson bracket for a function $f$ and a form $\a$ is,
$$\{f,\a\}=\pi^{ij}\d_if\nabla_{e_j}\a $$ This form of the Poisson bracket
agrees with the bracket put forth by \cite{ch} and \cite{hm}. The
action of the connection on a basis 1-form may be expressed with the
connection coefficients $\Gamma^i_{jk}$:
\begin{equation}\nabla_{e_i}dx^k \equiv -\Gamma_{ij}^k dx^j.\end{equation}
In general, this connection has torsion, so two connections can be
defined from the same connection coefficients:
\align{
\Gamma_i^k &= dx^\ell\Gamma_{\ell i}^k &
\t{\Gamma}_i^k&= \Gamma_{i\ell}^kdx^\ell,
}
which act on 1-forms as: 
\align{
\nabla_{e_i} dx^k  &=-\Gamma_{ij}^kdx^j= -\t{\Gamma}_i^k &
\t{\nabla}_{e_i}dx^k &= -\t{\Gamma}_{ij}^kdx^j=  -\Gamma_{ji}^kdx^j= -\Gamma_j^k ,
}
where $\t{\Gamma}_{ij}^k=\Gamma_{ji}^k$ are the connection coefficients
associated with $\t{\nabla}$. Notation is discussed in more detail 
in Appendix \ref{notation}.

The graded differential Poisson algebra structure imposes several
conditions upon the connections that appear in this expression for the
Poisson bracket.  For example, for the Leibniz rule to hold, the
Poisson bivector $\pi^{ij}$ must be covariantly constant under the
connection $\t{\nabla}_i$.  The Leibniz rule requires that:
\align{
d\{x^i,x^j\} &= \{dx^i,x^j\}+\{x^i,dx^j\} \\
d\pi^{ij} &= \pi^{jm}\Gamma_{mk}^idx^k-\pi^{im}\Gamma_{mk}^jdx^k.
}
Using the definition of the Poisson bracket to calculate each of the quantities, we see that imposing the Leibniz rule requires the following condition on the connection,
\be
(\t{\nabla}_{e_k}\pi)^{ij} =\d_k\pi^{ij}+\Gamma_{mk}^i\pi^{mj}+\Gamma_{mk}^j\pi^{im}=0. \label{sc}
\ee
Thus $\pi^{ij}$ is covariantly constant under $\t{\nabla}_{e_k}$, and,
equivalently, $\omega_{ij}$ is covariantly constant under
$\t{\nabla}_{e_k}$.  We refer to $\t{\nabla}_{e_k}$ as a
symplectic connection\footnote{In the literature, symplectic
connections are usually taken to be torsionless, but the connection
defined here generally has torsion.}, because it annihilates the
symplectic 2-form.  One can use the symplectic condition (\ref{sc}) to
rewrite the Jacobi identity (\ref{jac}) in terms of the torsion:
$$\pi^{im}\pi^{j\ell}\(\Gamma_{\ell m}^k-\Gamma_{m\ell}^k\) +\pi^{jm}\pi^{k\ell}\(\Gamma_{\ell m}^i- \Gamma_{m\ell}^i\) +\pi^{km}\pi^{i\ell}\(\Gamma_{\ell m}^j- \Gamma_{m\ell}^j\)=0$$
Note that while this relation shows that a torsionless connection identitically satisfies the Jacobi identity, the Jacobi identity does not require the connection to be torsionless.  

\subsection{The Poisson bracket between two 1-forms}
To find the form of the Poisson bracket for two forms, we use the
Leibniz rule for basis 1-forms:
$$\{dx^i,dx^j\}=d\{x^i,dx^j\}.$$
The Poisson bracket for a function and a 1-form gives:
\align{
\{dx^i,dx^j\} &= d\{x^i,dx^j\} \\
&= d(-\pi^{i\ell}\Gamma_{\ell b}^jdx^b)\\
&= -\d_a\pi^{i\ell}\Gamma_{\ell b}^jdx^a dx^b-\pi^{i\ell}\d_a\Gamma_{\ell b}^jdx^a dx^b.
}The symplectic condition of $\t{\nabla}_{e_k}$ relates the derivative of $\pi^{ij}$ to the connection coefficients $\Gamma_{ij}^k$, which gives
\align{
\{dx^i,dx^j\}&= (\Gamma_{ma}^i\pi^{m\ell}+\Gamma_{ma}^\ell\pi^{im})\Gamma_{\ell b}^j dx^adx^b -\pi^{i\ell}\d_a\Gamma_{\ell b}^jdx^adx^b\\
&= -\pi^{im}(\d_a\Gamma_{mb}^j-\Gamma_{\ell b}^j\Gamma_{ma}^\ell)dx^adx^b+\pi^{m\ell}\Gamma_{ma}^i\Gamma_{\ell b}^j dx^adx^b.
} The term in the parentheses can be rewritten in terms of the curvature
$\t{R}_{mab}^j$ of the symplectic connection $\t{\nabla}_{e_k}$ using the
anti-symmetry of $dx^adx^b$:\footnote{Note that $\t{R}^{ij}_{ab}
\equiv \pi^{im}\t{R}_{mab}^j$, which we often refer to as the
curvature.  Our notation is discussed further in Appendix
\ref{notation}.}
$$-\pi^{im}\cdot \frac12(\d_a\Gamma_{mb}^j-\d_b\Gamma_{ma}^j+\Gamma_{\ell a}^j\Gamma_{mb}^\ell-\Gamma_{\ell b}^j\Gamma_{ma}^\ell)dx^adx^b =  -\frac12 \pi^{im}\t{R}^{j}_{mab}dx^adx^b \equiv -\frac12 \t{R}^{ij}_{ab}dx^adx^b$$

The Poisson bracket between two basis 1-forms is then:
$$\{dx^i,dx^j\} =\pi^{m\ell}\Gamma_{ma}^i\Gamma_{\ell b}^j dx^adx^b-\frac12 \t{R}^{ij}_{ab}dx^adx^b.$$
To find the expression in terms of two arbitrary 1-forms, not just in terms of the basis $dx^i$, consider two 1-forms $\a,\b$ and use the product rule:
\align{
\{\a,\b\} &= \{\a_idx^i,\b_jdx^j\} \\
&= \{\a_i,\b_j\}dx^idx^j+dx^i\{\a_i,dx^j\}\beta_j+\a_i\{dx^i,\beta_j\}dx^j+\a_i\{dx^i,dx^j\}\beta_j \\
&= \pi^{mn}(\d_m\a_i -\Gamma_{mi}^\ell\a_\ell)dx^i(\d_n\b_j-\Gamma_{nj}^k\b_k)dx^j-\frac12 \a_i\b_j \t{R}^{ij}_{ab}dx^adx^b.
}After defining the 2-form
$$\t{R}^{ij}\equiv \frac12\t{R}^{ij}_{ab}dx^a dx^b,$$ 
the last line gives:
$$\{\a,\b\} = \pi^{mn}\nabla_{e_m}\a\nabla_{e_n}\b -\t{R}^{ij}\a_i\b_j.$$
For later convenience, let us also introduce the interior product, $i_{e_m}$\footnote{The interior product maps $p$-forms into
$(p-1)$-forms.  Our conventions for the interior product are discussed in Appendix \ref{notation}.}, and rewrite the bracket between two 1-forms as:
$$\{\a,\b\} = \pi^{mn}\nabla_{e_m}\a \nabla_{e_n} \b - \t{R}^{mn}(i_{e_m}\a)(i_{e_n}\b)
\qquad\qquad |\a|= |\b|=1.$$

An important property of the curvature tensor $\t{R}^{ij}$  is its
symmetry in the upper two indices. The graded symmetry of the
Poisson bracket requires $\{dx^i,dx^j\} = \{dx^j,dx^i\}$.  Using the Poisson bracket to calculate both sides, one finds:
$$\pi^{m\ell} \Gamma^i_{ma}\Gamma^j_{\ell b} dx^adx^b-\t{R}^{ij} = \pi^{m\ell} \Gamma^j_{ma}\Gamma^i_{\ell b} dx^adx^b-\t{R}^{ji}.$$
The ``$\Gamma^i\Gamma^j$'' term is manifestly symmetric in ($i,j$), from which it follows that $\t{R}^{ij}$ ought to be symmetric in the upper two indices.
We will soon show that $\t{R}^{ij}$ is the curvature of a symplectic (Poisson)
connection; that is $(\t{\nabla}_{e_i}\pi)^{jk}=0$. Like in Riemannian 
geometry where preserving the metric causes these indices to be 
anti-symmetric, preserving the symplectic 2-form\footnote{Or equivalently, the
Poisson bivector.}
causes these indices to be symmetric. This calculation is performed in
the appendix.

\subsection{The Poisson bracket between two arbitrary forms \label{p_b}}

It is straightforward to generalize the Poisson bracket for two
1-forms to forms of arbitrary degree.  We take the following ansatz
for the Poisson bracket, where now $\a, \b$ are forms of arbitrary
degree:
\begin{equation}\{\a,\b\} = \pi^{mn}\nabla_{e_m}\a\nabla_{e_n}\b +(-1)^{|\a|}\t{R}^{mn}(i_{e_m}\a)(i_{e_n}\b). \label{Ansatz}
\end{equation}
As shown in Appendix \ref{Ver1}, requiring (\ref{Ansatz}) to satisfy
the properties of the graded differential Poisson bracket places the
following conditions, also obtained in \cite{ch} and \cite{hm}, on the
connection coefficients $\Gamma^i_{jk}$:

\begin{enumerate}
\item  $\t{\nabla}_{e_\ell}$ is symplectic:
\be (\t{\nabla}_{e_\ell}\pi)^{ij}=0 \label{symp} \ee
\item $\pi^{ij}$ satisfies the Jacobi Identity: 
\begin{equation}\label{jacobi_pi}(\pi^{ab}\d_b\pi^{mn}+\pi^{mb}\d_b \pi^{na}+\pi^{nb}\d_b\pi^{am})=0\end{equation}
\item The connection $\nabla_{e_i}$ has vanishing curvature: 
\begin{equation}\label{zero_curv}[\nabla_{e_m},\nabla_{e_n}]\a=0 \end{equation}
\item The curvature $\t{R}^{mn}$ is covariantly constant under $\nabla_{e_a}$: 
\begin{equation}\label{cov_const}(\nabla_{e_a}\t{R})^{mn}=0\end{equation}
\item $\t{R}^{mn}$ satisfies\footnote{In tensor product notation, the
classical Yang-Baxter (CYB) equation is given by $\[r_{12},r_{13}\] +
\[r_{12},r_{23}\]+ \[r_{13},r_{23}\]= 0$, where $\[, \]$ is the matrix
commutator \cite{jf}.  Taking into account that $\t{R}^{ij}_{ab}$ is
symmetric in the two upper indices and antisymmetric in the two lower
indices, one can verify that (\ref{yang_baxter}) implies that
$\t{R}^{ij}_{ab}$ satisfies the CYB equation.}:

\begin{equation}\label{yang_baxter}\t{R}^{ab}(i_{e_b}\t{R}^{mn})+\t{R}^{mb}(i_{e_b}\t{R}^{na})+\t{R}^{nb}(i_{e_b}\t{R}^{am})=0 \end{equation}
\end{enumerate}
Equation \ref{symp} comes from requiring the Leibniz rule hold, and Equations \ref{jacobi_pi}-\ref{yang_baxter} come from requiring that the Jacobi identity hold.  These conditions, however, are not independent.  Consider the following relation:
\begin{multline*}
d\( \{x^i,\{dx^j,dx^k\}\} + \{dx^j,\{dx^k,x^i\}\}-\{dx^k,\{x^i,dx^j\}\}\) \\
= \{dx^i,\{dx^j,dx^k\}\} + \{dx^j,\{dx^k,dx^i\}\}-\{dx^k,\{dx^i,dx^j\}\}
\end{multline*}
After some reorganization, this equation relates (\ref{yang_baxter}) to the other conditions:
{\small
\begin{eqnarray} 
\lefteqn{ \t{R}^{ib}\(i_{e_b}\t{R}^{jk}\)+ \t{R}^{jb}\(i_{e_b}\t{R}^{ki}\)+ \t{R}^{kb}\(i_{e_b}\t{R}^{ij}\) =} \nonumber \\
&&\phantom{XXX}- \( \pi^{ab}\d_b \pi^{mn} + \pi^{mb}\d_b \pi^{na} + \pi^{nb}\d_b \pi^{am}\) \nabla_{e_a} dx^i\nabla_{e_m} dx^j\nabla_{e_n} dx^k \nonumber\\
&&\phantom{XXX} - \pi^{ab}\pi^{mn}\( \[\nabla_{e_a}, \nabla_{e_m}\]dx^i \nabla_{e_b} dx^j\nabla_{e_n} dx^k + \nabla_{e_n} dx^i\[\nabla_{e_a}, \nabla_{e_m}\]dx^j\nabla_{e_b} dx^k + \nabla_{e_b} dx^i\nabla_{e_n} dx^k\[\nabla_{e_a}, \nabla_{e_m}\]dx^k\) \nonumber \\
&&\phantom{XXX} + \pi^{ab}\(\nabla_{e_b} \t{R}\)^{jk}\nabla_{e_a} dx^i+ \pi^{ab}\(\nabla_{e_b} \t{R}\)^{ki}\nabla_{e_a} dx^j+ \pi^{ab}\(\nabla_{e_b} \t{R}\)^{ij}\nabla_{e_a} dx^k \nonumber \\
&&\phantom{XXX}+ d\bigg( \( \pi^{ib}\d_b \pi^{mn} + \pi^{mb}\d_b \pi^{ni} + \pi^{nb}\d_b \pi^{im} \) \nabla_{e_m} dx^j\nabla_{e_n} dx^k\bigg) \nonumber \\
&&\phantom{XXX}+ d\bigg( \pi^{ab}\pi^{mi}\( \[\nabla_{e_a}, \nabla_{e_m}\]dx^i \nabla_{e_b} dx^k + \nabla_{e_b} dx^j\[\nabla_{e_a}, \nabla_{e_m}\]dx^k\)\bigg) \nonumber \\
&&\phantom{XXX} -d\bigg(\pi^{ib}\(\nabla_{e_b}\t{R}\)^{jk}\bigg) + . . . \label{eww}
\end{eqnarray}
}where the ellipses represents terms proportional to $(\d_k
\pi^{ij}+\pi^{ia} \Gamma_{ak}^j +\pi^{aj} \Gamma_{ak}^i)$ and
derivatives thereof.  We have not written these terms since they
vanish using the symplectic condition (\ref{symp}).  Once one requires
(\ref{jacobi_pi}), (\ref{zero_curv}), and (\ref{cov_const}), relation
(\ref{eww}) implies:

$$\t{R}^{ib}\(i_{e_b}\t{R}^{jk}\)+ \t{R}^{jb}\(i_{e_b}\t{R}^{ki}\)+
\t{R}^{kb}\(i_{e_b}\t{R}^{ij}\)=0.$$
This shows that (\ref{yang_baxter}) is implied by the other
constraints.  Similar relations hold between Equations
\ref{jacobi_pi}-\ref{cov_const}.  However, none of these constraints
can be completely expressed in terms of the others.

If a connection exists that satisfies all of these properties on
the manifold, we have given expressions for the Poisson bracket between two
arbitrary differential forms. This bracket is the only possible bracket between
differential forms on a symplectic manifold.

\section{The Star Product \label{secstar}}
In deformation quantization, star products replace the point-wise
multiplication of functions on a Poisson manifold. For an overview of
the subject, see \cite{f} and \cite{ds}. The space of smooth
functions, $C^\infty(M)$, is replaced with
$C^\infty(M,*)[[\hbar]]$. As a set, $C^\infty(M,*)[[\hbar]]$ consists
of formal (infinite) polynomials in $\hbar$ with the coefficients
belonging to $C^\infty(M)$:
$$f(x)=f_0(x)+ \sum_{n=1}^\infty \hbar^n f_n(x).$$ Functions in
$C^\infty(M,*)[[\hbar]]$ are multiplied with the star product.  In the
limit $\hbar\to 0$, the space $C^\infty(M,*)[[\hbar]]$ becomes
$C^\infty(M)$, and the star product reduces to the usual point-wise
multiplication of functions.  Kontsevich \cite{k} showed that the star
product exists for all Poisson manifolds and gave a procedure to
obtain the formula to all orders.  The star product is written
explicitly to $\cO(\hbar^2)$ in \cite{k}, while more recently
\cite{kv} gave the formula to $\mathcal{O}\(\hbar^4\)$.  To
$\mathcal{O}\(\hbar^2\)$, the star product is:
$$f*g= fg +\hbar \{f, g\}+ \hbar^2 \Bigg(\frac12
\pi^{ij}\pi^{mn}\d_i\d_mf\d_j\d_ng
+\frac13\pi^{ia}\d_a\pi^{mn}\(\d_i\d_mf\d_ng-\d_mf\d_i\d_n\g\)\Bigg).$$
Two star products, $*_1$ and $*_2$, are considered equivalent if there
exists a linear operator $T$ of the form:
$$T= id + \sum_{i=n}^\infty\hbar^n T_n$$
such that
$$T\(f*_2 g\)= T\(f\) *_1 T\(g\).$$

\subsection{The Star Product of Differential Forms}

To enlarge the definition of the star product to include differential
forms, we deform $\Omega^*(M)$ to $\Omega^*(M,*)[[\hbar]]$, which, as
a set, consists of formal power series in $\hbar$ with coeffiecients
in $\Omega^*(M)$:
$$\a = \a_0 + \sum_{n=1}^\infty \hbar^n \a_n.$$
For $\a, \b \in \Omega^0(M,*)[[\hbar]]=C^\infty(M,*)[[\hbar]]$,  we require that
the star product of forms coincides with the usual star product 
of functions.  A product of differential forms
in $\Omega^*(M,*)[[\hbar]]$ is a ``star product'' if it satisfies the following properties for $ \a, \b, \g \in\Omega^*(M,*)[[\hbar]]$:
\begin{enumerate}
\item The product takes the form: 
$$\a*b=\a\b + \sum_{n=1}^\infty\hbar^n C_n(\a, \b)$$
where the $C_n$ are bilinear differential operators. \label{sum}
\item The product is associative: $\(\a*\b\)*\g = \a*\(\b*\g\)$ \label{star_assoc}
\item The order $\hbar$ term is Poisson bracket for forms: $C_1(\a, \b)=\{\a, \b\}$\label{star_pb}
\item The constant function, $1$, is the identity: $1*\a = \a*1 = \a$
\item The terms $C_n$ have a generalized Moyal symmetry: $$C_n(\a,\b)=(-1)^{|\a||\b|+n}C_n(\b,\a). \label{moyal}$$
\end{enumerate}  
We want to find a star product which satisifies Properties \ref{sum}-\ref{moyal} to
$\mathcal{O}\(\hbar^2\)$ for two arbitrary forms $\a$ and $\b$.  We propose the following product:
$$\a*\b = \a \b +\hbar C_1(\a,\b)+\hbar^2\[C^\nabla_2(\a,\b)+C^{\t{R}}_2(\a,\b)\]+\dots$$
where we have:
\begin{eqnarray}
C_1(\a,\b)&\equiv& \{\a,\b\} = \pi^{mn}\nabla_{e_m}\a\nabla_{e_n}\b +(-1)^{|\a|}\t{R}^{mn}(i_{e_m}\a)(i_{e_n}\b) \nonumber \\
C_2^\nabla(\a,\b) &=& \frac12 \pi^{ij}\pi^{mn}\nabla_{e_i}\nabla_{e_m}\a\nabla_{e_j}\nabla_{e_n}\b 
+\frac13\pi^{ia}\d_a\pi^{mn}\(\nabla_{e_i}\nabla_{e_m}\a\nabla_{e_n}\b-\nabla_{e_m}\a\nabla_{e_i}\nabla_{e_n}\b\)  \nonumber \\
C_2^{\t{R}}(\a,\b) &= &-\frac12 \t{R}^{ij}\t{R}^{mn}\[(i_{e_i}i_{e_m}\a)(i_{e_j}i_{e_n}\b)\] -\frac13 \t{R}^{i\ell}(i_{e_\ell} \t{R}^{mn})\bigg[(-1)^{|\a|}(i_{e_i}i_{e_m}\a)(i_{e_n}\b)+(i_{e_m}\a)(i_{e_i}i_{e_n}\b)\bigg]  \nonumber \\
&&+(-1)^{|\a|} \pi^{ij}\t{R}^{mn}(i_{e_m}\nabla_{e_i}\a)(i_{e_n}\nabla_{e_j}\b).  \label{star_ansatz}
\end{eqnarray}
We write the $\mathcal{O}(\hbar^2)$ term in this way for later convenience.  

Properties \ref{sum} and \ref{star_pb}-\ref{moyal} are manifestly
satisfied by (\ref{star_ansatz}); however we leave the calculation
that shows (\ref{star_ansatz}) is associative (Property
\ref{star_assoc}) to Appendix \ref{Ver2}, because the calculation is
quite involved.  However, constructing (\ref{star_ansatz}) for the
product to this order was relatively straightfoward.  The condition of
associativity at order $\hbar^2$ requires that for $f$ a function and
$\a,\b$ forms: 
\be 
f C_2(\a,\b)-C_2(f \a,\b)+C_2(f,\a\b)-C_2(f,\a)\b =
\{\{f,\a\},\b\} - \{f,\{\a,\b\}\}. \label{o2} 
\ee 
The right-hand side of (\ref{o2}) is calculable using the Poisson
bracket (\ref{Ansatz}).  The star product between a function and a
form to any order can be obtained by replacing ordinary derivatives in
the star product of functions with covariant derivatives.  This
prescription gives $C_2(f,\a\b)$ and $C_2(f,\a)$ in (\ref{o2}), so
constructing the ansatz for $C_2(\a,\b)$ simplifies considerably.
Using the same methods, constructing an ansatz for the star product to
$\cO(\hbar^3)$ is similarly straightforward, although verifying
the associativity of the product is laborous.

\subsection{Changing Coordinates}

As can easily be seen from the form of the star product, it behaves
very badly under a change in coordinates. Say we start with coordinates
$x^i$ and we want to switch to coordinates $y^\mu$. First we write the
$x^i$ as functions of $y^\mu$, e.g. $x^i(y^\mu)$. Then, we see how the 
basis vectors and basis 1-forms behave under this change of coordinates:
$$e_i = \frac{\d y^\mu}{\d x^i}e_\mu\qquad\qquad \text{and}\qquad\qquad 
dx^i = \frac{\d x^i}{\d y^\mu}dy^\mu$$
Where $\d y^\mu/\d x^i$ is the matrix inverse of $\d x^i/\d y^\mu$. We will
also use the fact that:
$$\frac{\d}{\d y^\mu} \(\frac{\d y^\lambda}{\d x^i}\) = -\frac{\d^2x^j}{\d y^\mu \d y^\nu}\frac{\d y^\nu}{\d x^i}\frac{\d y^\lambda}{\d x^j}.$$
When we plug these expressions into the above formula for the star product,
we get the same terms, except now in the $y$-coordinate basis, plus
additional terms. To be precise, after a coordinate change, we get:
\align{
\a *'\b &=\a\b + \hbar C_1(\a,\b) + \hbar^2 \[C_2^\nabla(\a,\b)+C_2^{\t R}(\a,\b)\]  \\
& \phantom{\a\b} +\frac{\hbar^2}{6}\bigg[-\frac{\d^2x^i}{\d y^\mu\d y^\nu}
\frac{\d y^\lambda}{\d x^i}\pi^{\mu\rho}\pi^{\nu \sigma} \(\nabla_{e_\rho}\nabla_{e_\sigma}\a\nabla_{e_\lambda}\b+\nabla_{e_\lambda}\a\nabla_{e_\mu}\nabla_{e_\nu}\b +
2\nabla_{e_\mu}\nabla_{e_\lambda}\a\nabla_{e_{\nu}}\b+2\nabla_{e_\mu}\a\nabla_{e_\nu}\nabla_{e_\lambda}\b\) \\
&\phantom{\a\b-\frac{\hbar^2}{6}\bigg[}+
\bigg(\frac72 \frac{\d^2x^i}{\d y^\mu\d y^\nu}\frac{\d^2 x^j}{\d y^\lambda \d y^\tau}\frac{\d y^\rho}{\d x^i}\frac{\d y^\sigma}{\d x^j}\pi^{\mu \lambda}\pi^{\nu\tau}+2\frac{\d^2 x^i}{\d y^\mu\d y^\nu}\frac{\d^2 x^j}{\d y^\lambda \d y^\tau}\frac{\d y^\nu}{\d x^j}\frac{\d y^\rho}{\d x^i}\pi^{\mu \lambda}\pi^{\tau \sigma}
 \\
&\phantom{=\a\b+ \phantom{\a\b-\frac{\hbar^2}{6}\bigg[}-
\bigg(\frac72 \frac{\d^2x^i}{\d y^\mu\d y^\nu}\frac{\d^2 x^j}{\d y^\lambda \d y^\tau}}+2\frac{\d^2x^i}{\d y^\mu\d y^\nu}\frac{\d y^\rho}{\d x^i} \pi^{\mu\lambda}\d_\lambda \pi^{\nu\sigma}  \bigg)
\(\nabla_{e_\rho}\a\nabla_{e_\sigma}\b + \nabla_{e_\sigma}\a \nabla_{e_\rho}\b\) \bigg]
}
Where $C_1(\a,\b),~C_2^\nabla(\a,\b)$, and $C_2^{\t R}(\a,\b)$ take the same
form, except now in the $y^\mu$ coordinates; e.g. 
$C_1(\a,\b)=\pi^{\mu\nu}\nabla_{e_\mu}\a\nabla_{e_\nu}\b +(-1)^{|\a|}\t R^{\mu\nu}
(i_{e_\mu}\a)(i_{e_\nu}\b)$. As we can see, this is not the form we want for
the star product, and the difference appears at order $\hbar^2$. It turns out
that this formula for the star product is actually equivalent (in the 
sense discussed above) to the original formula for the star product.
To see this, we will need to construct a differential operator $T$ such that
$$T(\a * \b) = T(\a) *' T(\b)\qquad \Rightarrow \qquad \a * \b = T^{-1}\(T(\a)*'T(\b)\)$$
We need to make changes at order $\hbar^2$, so take $T$ to be of the form
$T=1+\hbar^2T_2+\dots$ where $T_2$ is the differential operator that
contributes to $T$ at order $\hbar^2$.
Then the formula above implies that:
$$\a * \b = \a *' \b - \hbar^2 T_2(\a\b) + \hbar^2 (T_2(\a))\b+\hbar^2\a(T_2(\b))+\cO(\hbar^3).$$
It turns that the operator $T_2$ required to eliminate the unwanted terms
is given by:
\align{
T_2(\a) &=\frac{1}{6}\bigg[-\frac{\d^2 x^i}{\d y^\mu \d y^\nu}\frac{\d y^\lambda}{\d x^i}\pi^{\mu\rho}\pi^{\nu\sigma}\nabla_{e_\lambda}\nabla_{e_\rho}\nabla_{e_\sigma}
+\frac72 \frac{\d^2 x^i}{\d y^\mu \d y^\nu}\frac{\d^2 x^j}{\d y^\lambda \d y^\tau}\frac{\d y^\rho}{\d x^i}\frac{\d y^\sigma}{\d x^j}\pi^{\mu\lambda}\pi^{\nu\tau}\nabla_{e_\rho}\nabla_{e_\sigma}  \\
&\phantom{\frac{\hbar^2}{6}\bigg[-\frac{\d^2 x^i}{\d y^\mu \d y^\nu}\frac{\d y^\lambda}{\d x^i}} +2 \frac{\d^2 x^i}{\d y^\mu\d y^\nu} \frac{\d^2 x^j}{\d y^\lambda \d y^\tau}\frac{\d y^\mu}{\d x^j}\frac{\d y^\rho}{\d x^i}\pi^{\nu\lambda}\pi^{\tau\sigma}\nabla_{e_\rho}\nabla_{e_\sigma}
+2 \frac{\d^2 x^i}{\d y^\mu\d y^\nu}\frac{\d y^\rho}{\d x^i}\pi^{\mu\lambda}\d_\lambda \pi^{\nu\sigma}\nabla_{e_\rho}\nabla_{e_\sigma} \bigg]\a
}
Though not very illuminating, it is comforting to know that even though the
product appears to be badly behaved under a change in coordinates, there
is a transformation that brings it back into the form we started out with.

\section{Discussion}

The graded differential Poisson algebra on a symplectic manifold was
initially introduced by \cite{ch} and \cite{hm}.  These papers contain
an explicit form of the bracket for 1-forms. In this note, we have
generalized the explicit form of the graded Poisson bracket for all
degrees of forms, and checked that the form satisfies all the
required properties.  We found that for the Poisson bracket to satisfy
the properties required of a graded differential Poisson algebra, the
symplectic manifold must satisfy three new conditions\footnote{We already
have $\pi^{ij}= -\pi^{ij}$ and
$\pi^{im}\d_m\pi^{jk}+\pi^{jm}\d_m\pi^{ki}+\pi^{km}\d_m\pi^{ij}=0$
from the Poisson algebra of functions.}, also found by  \cite{ch} and \cite{hm}:
\begin{enumerate}
\item $\t{\nabla}_{e_\ell}$  is symplectic: $\t{\nabla}_{e_\ell}\pi^{ij}=0$ \label{1}
\item $R^i_{mab}$ is flat: $[\nabla_{e_m},\nabla_{e_n}]\a=0$ \label{2}
\item $\t{R}^{ij}$ is covariantly constant: $\(\nabla_{e_k} \t{R}\)^{ij} = 0$  \label{3}
\end{enumerate}
These conditions come from requiring that the graded Poisson bracket
satisfy the Leibniz rule (\ref{1}) and the Jacobi identity
(\ref{2}-\ref{3}).  They are not independent.

We then use this graded differential Poisson bracket in our
presentation of a non-commutative deformation (the star product) of
the graded differential algebra of forms to $\mathcal{O}\(\hbar^2\)$.
The star product we present is:
$$\a*\b = \a \b +\hbar C_1(\a,\b)+\hbar^2C_2(\a,\b),$$
where we have:
\align{
C_1(\a,\b)&\equiv \{\a,\b\} = \pi^{mn}\nabla_{e_m}\a\nabla_{e_n}\b +(-1)^{|\a|}\t{R}^{mn}(i_{e_m}\a)(i_{e_n}\b)\\
C_2(\a,\b)&=C_2^\nabla(\a,\b)+ C_2^{\t{R}}(\a,\b) \\
&C_2^\nabla(\a,\b)= \frac12 \pi^{ij}\pi^{mn}\nabla_{e_i}\nabla_{e_m}\a\nabla_{e_j}\nabla_{e_n}\b 
+\frac13\pi^{ia}\d_a\pi^{mn}\(\nabla_{e_i}\nabla_{e_m}\a\nabla_{e_n}\b-\nabla_{e_m}\a\nabla_{e_i}\nabla_{e_n}\b\) \\
&C_2^{\t{R}}(\a,\b) = -\frac12 \t{R}^{ij}\t{R}^{mn}\[(i_{e_i}i_{e_m}\a)(i_{e_j}i_{e_n}\b)\] -\frac13 \t{R}^{i\ell}(i_{e_\ell} \t{R}^{mn})\bigg[(-1)^{|\a|}(i_{e_i}i_{e_m}\a)(i_{e_n}\b)+(i_{e_m}\a)(i_{e_i}i_{e_n}\b)\bigg] \\
&+(-1)^{|\a|} \pi^{ij}\t{R}^{mn}(i_{e_m}\nabla_{e_i}\a)(i_{e_n}\nabla_{e_j}\b),
}
and we verify that this star product satisifes all the necessary
properties, such as associativity, to $\mathcal{O}\(\hbar^2\)$.

The results of this paper may be generalized in two directions.  Using
the method outlined in Section \ref{secstar}, it should be possible
(although tedious) to find the star product between differential forms
to $\cO(\hbar^3)$.

It may also be possible to generalize the star product to Poisson
manifolds, where $\pi^{ij}$ is not invertible. Many of the arguments
go through unchanged. However, the form of the star product between a
function and a form is not guaranteed in this case, and the conditions
on the connection $\Gamma_{ij}^k$ are more complicated (e.g.
$\pi^{ia}\pi^{jb}R_{mab}^k=0$ instead of $R_{mab}^k=0$).

Finally, we can apply the star product between differential forms to
physics. Applications include considering more general star products
in deformation quantization, to studying gauge theories on
noncommutative spaces, to generalizing the Seiberg-Witten map
\cite{sw}.

\begin{acknowledgements}
I would like to thank Bruno Zumino for helpful advise and for pointing
me in this direction. I would like to thank Shannon McCurdy for checking
the calculations contained in this paper and for helping to write a draft
of this paper.
This work was supported in part by the Director, Office of Science,
Office of High Energy and Nuclear Physics, Division of High Energy
Physics of the U.S. Department of Energy under Contract
No. DE-AC03-76SF00098, in part by the National Science Foundation under
grant PHY-0457315.\end{acknowledgements}

\appendix

\section{Notation \label{notation}}

\subsection{Basis Vectors and Basis 1-forms}

Vectors can be thought of as linear differential operators acting on
smooth functions over a manifold. In a coordinate chart spanned by 
$x^1,\dots,x^n$, there is a natural set of basis vectors, $e_i$, given by:
$$e_i(f) \equiv \frac{\d f}{\d x^i}~\(=\d_if\)$$
The 1-forms are dual to these vectors and can be thought of as linear
operators turning a vector field into a function. They are spanned by
basis 1-forms $dx^i$ that are defined by:
$$dx^i(e_j) = \delta^i_{\phantom{i}j}$$

\subsection{The Interior Product}

The interior product is a map $i_V:~\Omega^n(M)\to
\Omega^{n-1}(M)$. For an $n$-form the interior product is given by the
formula:
$$(i_V\omega)(X_1,\dots,X_{n-1})=\omega(V,X_1,\dots,X_{n-1})$$
We are concerned with the interior product of a basis vector acting on
a differential form in this paper. We will typically take the interior
product with respect to the basis vectors $e_i$.

Using this notation, here are a few of the properties of the interior product:
\align{
\a &= \frac{1}{p!}\a_{i_1\dots i_p}dx^{i_1}\dots dx^{i_p} \\
i_{e_m} \a & = \frac{1}{(p-1)!}\a_{mi_2\dots i_p}dx^{i_2}\dots dx^{i_p} \\
i_{e_m} (\a\wedge \b) &= (i_{e_m}\a)\wedge \b + (-1)^{|\a|}\a\wedge (i_{e_m} \b) \\
i_{e_m}i_{e_n}\a &= -i_{e_n}i_{e_m}\a
}
In essence, the interior product does the following:
$$i_{e_m} dx^k = \delta^k_{\phantom{k}m}.$$
It maps 1-forms to numbers and it satisfies the graded product rule as
given above. This is the natural tool to use when describing the
Poisson bracket of higher degree forms.

\subsection{The connections $\nabla_{e_i},\t{\nabla}_{e_i}$}

We use the convention that covariant derivatives only act nontrivially
on basis 1-forms, $dx^i$ and basis vectors $e_i$.
They act like partial derivatives on the coefficients of 1-forms (and
other tensors), regardless of the index structure. For instance:
$$\nabla_{e_i} \a_k = \d_i\a_k,\qquad \text{while}\qquad \nabla_{e_i} dx^k = -\Gamma_{ij}^k dx^j.$$
Combined, these give the expected result,
$$\nabla_{e_i} (\a_k dx^k) = (\d_i \a_k)dx^k + \a_k (-\Gamma_{ij}^kdx^j) = (\d_i \a_j -\Gamma_{ij}^k\a_k)dx^j.$$
Note that the details of the calculation differ from the convention
for covariant derivatives widely used by physicists. What physicists 
typically call the ``covariant derivative'' is reproduced by the so-called
covariant differential:
$$\nabla \a = (\nabla_{e_i}\a)\otimes dx^i.$$ We have explicitly written
the full vector $e_i$ as the argument of the covariant derivative, 
$\nabla_{e_i}$, rather than using the conventional $\nabla_i$ to avoid
any potential confusion.

As stated in the text, we have defined the two connections as:
$$ \nabla_{e_i} dx^k =-\Gamma_{ij}^kdx^j \phantom{XXX}  \t{\nabla}_{e_i}dx^k = -\t{\Gamma}_{ij}^kdx^j= -\Gamma_{ji}^kdx^j.$$
The curvature for these two connections are given as:
\align{
[\nabla_{e_a},\nabla_{e_b}]dx^i&=R^i_{mab}dx^m = 
\(\d_a\Gamma^i_{bm}-\d_b\Gamma^i_{am}+\Gamma^i_{a \ell}\Gamma^\ell_{bm}-\Gamma^i_{b \ell}\Gamma^\ell_{am}\)dx^m \\
[\t{\nabla}_{e_a},\t{\nabla}_{e_b}]dx^i&=\t{R}^i_{mab}dx^m = 
\(\d_a\Gamma^i_{mb}-\d_b\Gamma^i_{ma}+\Gamma^i_{\ell a}\Gamma^\ell_{mb}-\Gamma^i_{\ell b}\Gamma^\ell_{ma}\)dx^m
} 
Without torsion, these two expressions would be equal, but since the
torsion is non-zero, these two curvatures are different from each
other.  In particular, one of these curvatures is zero while the other
is non-vanishing.

We also define
$$\t{R}^{ij}_{ab} \equiv \pi^{im}\t{R}_{mab}^j$$
where we raised one of the lower indices of the curvature tensors with
the Poisson bivector $\pi$.  We often also refer to $\t{R}^{ij}_{ab}$
as the curvature.

When the covariant derivative acts on mixed tensors, we use
parentheses as a shorthand notation. For example,
$(\nabla_{e_c}\t{R})^{ij}$ is the $(i, j)$ component of
$$\nabla_{e_c}\t{R}=\frac12\bigg[ \d_c \t{R}^{ij}_{ab} + \Gamma_{c\ell}^i\t{R}^{\ell j}_{ab} + \Gamma_{c\ell}^j\t{R}^{i\ell}_{ab} -\Gamma^\ell_{ca}\t{R}^{ij}_{\ell b} -\Gamma^\ell_{cb}\t{R}^{ij}_{a\ell}\bigg] dx^a\wedge dx^b\otimes\d_i\otimes\d_j.$$
In particular, we have the following identity:
$$\nabla_{e_c} \[\t{R}^{mn}(i_{e_m}\a)(i_{e_n}\b)\] = (\nabla_{e_c}\t{R})^{mn}(i_{e_m}\a)(i_{e_n}\b) + \t{R}^{mn}(i_{e_m}\nabla_{e_c}\a)(i_{e_n}\b) + \t{R}^{mn}(i_{e_m}\a)(i_{e_n}\nabla_{e_c}\b)$$

\subsection{Showing $\t{R}^{ij}$ is Symmetric in the upper two indices}

First, remember that the connection $\t{\nabla}_{e_i}$ preserves the 
Poisson bivector:
$$(\t{\nabla}_{e_i}\pi)^{jk}=\d_i \pi^{jk}+\t{\Gamma}_{i\ell}^j\pi^{\ell k}+\t{\Gamma}_{i\ell}^k\pi^{j\ell}=0$$
Next, we manipulate the curvature tensor, using the above relation:
\align{
\t{R}^{ij}_{ab} \equiv \pi^{i\ell}\t{R}_{\phantom{j}\ell ab}^j &= \pi^{i\ell} \d_a \t{\Gamma}_{b\ell}^j + \pi^{i\ell}\t{\Gamma}_{am}^j\t{\Gamma}_{b\ell}^m -~(a\leftrightarrow b) \\
&= \[\d_a\(\pi^{i\ell}\t{\Gamma}_{b\ell}^j\)-\d_a\pi^{i\ell}\t\Gamma_{b\ell}^j\] +\t\Gamma_{am}^j\(\pi^{i\ell}\t\Gamma_{b\ell}^m\) -~(a\leftrightarrow b)\\
&= \d_a\(-\d_b\pi^{ij}-\pi^{\ell j}\t\Gamma_{b\ell}^i\) -\d_a\pi^{i\ell}\t\Gamma_{b\ell}^j+\t\Gamma_{am}^j \(-\d_b\pi^{im}-\pi^{\ell m}\t\Gamma_{b\ell}^i\) -~(a\leftrightarrow b)
}
Notice that $\d_a\d_b \pi^{ij}$ and $(\d_a\pi^{i\ell}\t\Gamma_{b\ell}^j 
+ \d_b\pi^{im}\t\Gamma_{am}^j)$ vanish when we anti-symmetrize the $a,b$
indices. Continuing,
\align{
\t R^{ij}_{ab}&= -\d_a\pi^{\ell j}\t\Gamma_{b\ell}^i - \pi^{\ell j}\d_a\t\Gamma_{b\ell}^j -\(\pi^{\ell m}\t\Gamma_{am}^j\)\t\Gamma_{b\ell}^i -~(a\leftrightarrow b) \\
&= -\d_a\pi^{\ell j}\t\Gamma_{b\ell}^i - \pi^{\ell j}\d_a\t\Gamma_{b\ell}^j -\(-\d_a\pi^{\ell j}-\pi^{mj}\t\Gamma_{am}^\ell \)\t\Gamma_{b\ell}^i -~(a\leftrightarrow b)\\
&=-\pi^{mj}\(\d_a\t\Gamma_{bm}-\t\Gamma_{b\ell}^i\t\Gamma_{am}^\ell\)-~(a \leftrightarrow b) \\
&= +\pi^{jm}\t R_{\phantom{i}mab}^i = \t R_{ab}^{ji}
}
The change in sign comes from switching the indices in $\pi^{mj}$, which is
anti-symmetric. This is identical to the proof that the Levi-Civita connection
is anti-symmetric in the same indices, except for this last step where instead
of having a symmetric metric, we have an anti-symmetric Poisson bivector.

\section{Graded differential Poisson bracket properties  \label{Ver1}}
We verify the ansatz of Equation (\ref{Ansatz}) satisfies the properties of a graded differential Poisson bracket, Equation (\ref{BD}-\ref{Graded}).
To save space on the page, we will abbreviate $\nabla_{e_m}$ as $\nabla_m$ and
we will abbreviate $i_{e_m}$ as $i_m$. Otherwise, equations would not fit on
the page.
\subsection{Bracket degree, $|\{\a,\b\}|=|\a|+|\b|$:}

The covariant derivative does not change the degree
of the differential form, $|\nabla_a\a|=|\a|$, while the interior product does,
$|i_m\a|=|\a|-1$.  The first term of the Poisson bracket has the proper degree:
$$|\nabla_a\a\nabla_b\b| = |\nabla_a\a||\nabla_b\b|=|\a|+|\b|$$
as does the second term:
$$|\t{R}^{mn}(i_m\a) (i_n\b)|=|\t{R}^{mn}|+|i_m\a|+|i_n\b| = 2+(|\a|-1)+(|\b|-1)=|\a|+|\b|.$$

\subsection{Graded Symmetry, $\{\a,\b\} = (-1)^{|\a||\b|-1}\{\b,\a\}$:}
The first term in the Poisson bracket has the proper graded symmetry by inspection:
$$\pi^{mn}\nabla_m\a\nabla_n\b = \pi^{mn} (-1)^{|\a||\b|}\nabla_n\b\nabla_m\a = (-1)^{(|\a||\b|-1)}~\pi^{nm}\nabla_n\b\nabla_m\a.$$

The additional minus sign comes from the antisymmetry of the Poisson
bivector: $\pi^{nm}=-\pi^{mn}$. The second term in the bracket also
has the proper grading:
\align{
(-1)^{|\a|}\t{R}^{mn}(i_m\a)(i_n\b) &=(-1)^{|\a|}\t{R}^{mn}(-1)^{(|\a|-1)(|\b|-1)}(i_n\b)(i_m\a)\\ 
&=(-1)^{|\a||\b|+|\b|-1}\t{R}^{mn}(i_n\b)(i_m\a) \\
&= (-1)^{|\a||\b|-1}\[ (-1)^{|\b|}\t{R}^{nm}(i_n\b)(i_m\a)\].
}
In the last line, we use the fact that $\t{R}^{mn}$ is symmetric in
the upper two indices. 

\subsection{Graded Product, $\{\a\wedge \b,\g\} = \a\wedge\{\b,\g\} + (-1)^{|\b||\g|}\{\a,\g\}\wedge \b$:}
To show that the Poisson bracket satisfies the graded product rule, consider:
{\small
\align{
\{\a\b,\g\} &= \pi^{mn}\nabla_m(\a\b)\nabla_n\g + (-1)^{|\a|+|\b|}\t{R}^{mn}i_m(\a\b)i_n\g \\
&=\pi^{mn}\[(\nabla_m\a)\b + \a(\nabla_m\b)\]\nabla_n\g+(-1)^{|\a|+|\b|}\t{R}^{mn}\[(i_m\a)\b + (-1)^{|\a|}\a (i_m\b)\]i_n \g \\
&= \a\[\pi^{mn}\nabla_m\b\nabla_n\g +(-1)^{|\b|}\t{R}^{mn}i_m\b i_n\g\] + (-1)^{|\b||\g|}\[\pi^{mn}\nabla_m\a\nabla_n\g +(-1)^{|\a|}\t{R}^{mn}i_m\a i_n\g \] \b \\
&= \a\{\b,\g\} +(-1)^{|\b||\g|}\{\a,\g\}\b,
}
}
where we use $\b\g=(-1)^{|\b||\g|}\g\b$ and $|i_m\b|=|\b|-1$ to go
from the second line to the third line.

\subsection{Leibniz Rule, $d \{\a, \b\}= \{d\a, \b\} + (-1)^{|\a|}\{\a,
d\b\}$:}
To verify that our bracket satisfies the Liebniz rule, we take the
differential of our Poisson bracket directly:
\align{
  d\{\a, \b\}&= d\[\pi^{ij}\nabla_i\a \nabla_j\b +
(-1)^{|\a|}\t{R}^{ij}\(i_i\a\) \(i_j \b\)\] \\
&= d\pi^{ij} \nabla_i\a \nabla_j\b + \pi^{ij} \[ \(d \nabla_i\a\)
\nabla_j\b + (-1)^{|\a|}\nabla_i\a \(d\nabla_j\b\) \] \\
&\phantom{==} + (-1)^{|\a|}\[d\t{R}^{ij} \(i_i\a\) \(i_j \b\) + \t{R}^{ij}
\( \(d i_i\a\) \(i_j \b\) + (-1)^{|\a|-1}\(i_i\a\) \(d i_j \b \)  \) \].
}  Note that $ \mathcal{L}_j= d i_j + i_j d$, where  $\mathcal{L}$ is the
Lie derivative, and $\mathcal{L}_j\g= \d_j\g$ for a form $\g$
\footnote{We take $\d_k\a$ to mean: $$\d_k\a=\frac{1}{p!}(\d_k\a_{i_i\dots
i_p})dx^{i_1}\cdots dx^{i_p}$$}.
Also, the covariant derivative can be rewritten as $\nabla_i \g=
\d_i\g -\t{\Gamma}_i^k \(i_k\g\)$.  Combining these two observations
results in a useful identity:
$$ \pi^{ij}d\nabla_j \g =\pi^{ij}\[ \nabla_j d\g + \t{\Gamma}_j^k \d_k \g
- d\t{\Gamma}_j^k\(i_k\g\) \]
=  \pi^{ij} \nabla_j d\g+\pi^{ij}\t{\Gamma}_j^k\nabla_k \g
-\t{R}^{ij}\(i_j \g\) $$
Rewriting $d\{\a,\b\}$ using this identities,
{\small
\align{
 d \{\a, \b\}&= d\pi^{ij} \nabla_i\a \nabla_j\b + \(\pi^{ij}   \nabla_i
d\a + \pi^{ij} \t{\Gamma}_i^k \nabla_k\a + \t{R}^{ij}\(i_i\a \)\)
\nabla_j\b  + (-1)^{|\a|}\nabla_i\a\(  \pi^{ij} \nabla_j d\b +
\pi^{ij}\t{\Gamma}_j^k \(i_k \b\) - \t{R}^{ij}\(i_j\b\)\) \\
& \phantom{=}+ (-1)^{|\a|}\t{R}^{ij} \[ \( \d_i \a- \(i_i d\a\)\) \(i_j
\b\) + (-1)^{|\a|-1}\(i_i\a\) \( \d_j\b - i_j d\b\)  \] +
(-1)^{|\a|}d\t{R}^{ij} \(i_i\a\) \(i_j \b\)\\
&= \{d\a, \b\} + (-1)^{|\a|}\{\a, d\b\} + \(d\pi^{ij} +
\pi^{kj}\t{\Gamma}_k^i + \pi^{ik}\t{\Gamma}_k^j \) \nabla_i\a \nabla_j\b 
+ (-1)^{|\a|} \( d\t{R}^{ij}+ \t{R}^{ik}\t{\Gamma}_k^j +
\t{R}^{kj}\t{\Gamma}_k^i \)  \(i_i\a\) \(i_j \b\).
}
}Let us inspect $d\t{R}^{ij}+ \t{R}^{ik}\t{\Gamma}_k^j +
\t{R}^{kj}\t{\Gamma}_k^i$.  Recall that $\t{R}^{ij}= \pi^{ik}\(
d\t{\Gamma}_k^j + \t{\Gamma}_{\ell}^j \t{\Gamma}_{k}^{\ell}\)$.  Then, with
a bit of rewriting,
\align{
 d\t{R}^{ij}+ \t{R}^{ik}\t{\Gamma}_k^j + \t{R}^{kj}\t{\Gamma}_k^i&=
\(d\pi^{ik} +\pi^{i\ell} \t{\Gamma}_{\ell}^k +\pi^{\ell
k}\t{\Gamma}_{\ell}^i\)d\t{\Gamma}_k^j + \(d\pi^{ik} +\pi^{im}
\t{\Gamma}_{m}^k +\pi^{m k}\t{\Gamma}_{m}^i\)\t{\Gamma}_{\ell}^j
\t{\Gamma}_{k}^{\ell}\\
&= \(d\pi^{ik} +\pi^{i\ell} \t{\Gamma}_{\ell}^k +\pi^{\ell
k}\t{\Gamma}_{\ell}^i\)\t{R}_k^j.
}
Using this result, $d\{\a,\b\}$ becomes:
$$d\{\a, \b\}= \{d\a, \b\} + (-1)^{|\a|}\{\a, d\b\} + \(d\pi^{ij} +
\pi^{kj}\t{\Gamma}_k^i + \pi^{ik}\t{\Gamma}_k^j \) \bigg[ \nabla_i\a
\nabla_j\b + (-1)^{|\a|}\t{R}_j^m\(i_i\a\) \(i_m \b\)\bigg].$$
Thus, for the Poisson bracket to satisfy the Leibniz rule for all forms,
one must apply the constraint on the connection coefficients that $\Gamma^i_{jk}$ that $\t{\nabla_\ell}$ is symplectic (Equation \ref{sc}).  Then,
$$d\{\a, \b\}= \{d\a, \b\} + (-1)^{|\a|}\{\a, d\b\}.$$

\subsection{Graded Jacobi Identity, $\{\a,\{\b,\g\}\} + (-1)^{|\a|(|\b|+|\g|)} \{\b,\{\g,\a\}\}+(-1)^{(|\a|+|\b|)|\g|}\{\g,\{\a,\b\}\}=0$:}

Next, we check that the ansatz for the Poisson bracket satisfies the
graded Jacobi Identity.  We use the Poisson bracket to calculate
$\{\a,\{\b,\g\}\}$, and after simplifying, we have:
{\small
\begin{eqnarray*}
\lefteqn{\{\a,\{\b,\g\}\} = \pi^{ab}\d_b \pi^{mn}\nabla_a\a\nabla_m\b\nabla_n\g + \pi^{ab}\pi^{mn}\(\nabla_a\a\nabla_b\nabla_m\b\nabla_n\g +\nabla_a\a\nabla_m\b\nabla_b\nabla_n\g \)}&& \\
&& \phantom{\{\a,\{\b,\g\}\} =}+\pi^{ab}\t{R}^{mn} \Bigg[ (-1)^{|\b|}(\nabla_a\a) (i_m\nabla_b\b)( i_n\g) + (-1)^{|\b|}(\nabla_a\a) (i_m\b) (i_n \nabla_b\g)\\
&& \phantom{\{\a,\{\b,\g\}\} = +\pi^{ab}\t{R}^{mn} \Bigg[}+ (-1)^{|\a|+|\b|}(i_m\a) (i_n\nabla_a\b) (\nabla_b\g)+(-1)^{|\a|+|\b|}(i_m\a) (\nabla_a \b) (i_n\nabla_b \g) \Bigg] \\
&& \phantom{\{\a,\{\b,\g\}\} =}+(-1)^{|\a|+|\b|}\t{R}^{ab}\t{R}^{mn}\bigg[(i_n\a) (i_b i_m\b)( i_n\g) +(-1)^{|\b|-1}(i_a\a) (i_m\b) (i_b i_n\g)\bigg] \\
&& \phantom{\{\a,\{\b,\g\}\} =}+(-1)^{|\b|}\pi^{ab}(\nabla_b\t{R})^{mn}\nabla_a\a (i_m\b)(i_n\g)+(-1)^{|\b|-1}\t{R}^{ab} (i_b\t{R}^{mn})~(i_a\a) (i_m\b) (i_n\g).
\end{eqnarray*}
}Cycling through $\a,\b$, and $\g$ we find that $\big[\{a,\{\b,\g\}\}+(-1)^{|\a|(|\b|+|\g|)}\{\b,\{\g,\a\}\} + (-1)^{|\g|(|\a|+|\b|)}\{\g,\{\a,\b\}\}\big] $ does not automatically vanish.  Instead: 
{\small 
\begin{eqnarray*}
\lefteqn{\{a,\{\b,\g\}\}+(-1)^{|\a|(|\b|+|\g|)}\{\b,\{\g,\a\}\} + (-1)^{|\g|(|\a|+|\b|)}\{\g,\{\a,\b\}\} = }  \\
&& \phantom{\{a,\{\b,\g\}\}+} (\pi^{ab}\d_b\pi^{mn}+\pi^{mb}\d_b \pi^{na}+\pi^{nb}\d_b\pi^{am})\nabla_a\a\nabla_m\b \nabla_n\g \\
&& \phantom{\{a,\{\b,\g\}\}+}+\pi^{ab}\pi^{mn} \bigg[ \([\nabla_a,\nabla_m]\a\) \nabla_b\b\nabla_n\g + \nabla_n \a \([\nabla_a,\nabla_m]\b\) \nabla_b \g + \nabla_b\a\nabla_n\b \([\nabla_a,\nabla_m]\g\)\bigg] \\
&&\phantom{\{a,\{\b,\g\}\}+}+(-1)^{|\b|}\pi^{ab}(\nabla_b\t{R})^{mn}\bigg[\nabla_a\a(i_m\b)(i_n\g)+(-1)^{|\a|+1}(i_n\a)\nabla_a\b(i_m\g)+(-1)^{|\a|+|\b|}(i_m\a)(i_n\b)\nabla_a\g\bigg]\\
&& \phantom{\{a,\{\b,\g\}\}+}+(-1)^{|\b|-1}\bigg[ \t{R}^{ab}(i_b\t{R}^{mn})+\t{R}^{mb}(i_b\t{R}^{na})+\t{R}^{nb}(i_b\t{R}^{am})\bigg] (i_a\a) (i_m\b) (i_n\g).
\end{eqnarray*}
}This gives the condition that the bivector $\pi^{ij}$ is a Poisson bivector as well as several new conditions on the
connection coefficients $\Gamma_{ij}^k$.  In particular we find:
\begin{enumerate}
\item $\pi^{ij}$ satisfies the Jacobi Identity: 
\begin{equation*}(\pi^{ab}\d_b\pi^{mn}+\pi^{mb}\d_b \pi^{na}+\pi^{nb}\d_b\pi^{am})=0\end{equation*}
\item The connection $\nabla_i$ has vanishing curvature: 
\begin{equation*}[\nabla_m,\nabla_n]\a=0 \end{equation*}
\item The curvature $\t{R}^{mn}$ is covariantly constant under $\nabla_a$: 
\begin{equation*}(\nabla_a\t{R})^{mn}=0\end{equation*}
\item $\t{R}^{mn}$ satisfies: 
\begin{equation*}\t{R}^{ab}(i_b\t{R}^{mn})+\t{R}^{mb}(i_b\t{R}^{na})+\t{R}^{nb}(i_b\t{R}^{am})=0 \end{equation*}
\end{enumerate}

These conditions, however, are not independent. This is discussed in Section \ref{p_b}.

\section{Verifying the product ansatz is a Star Product  \label{Ver2}}

The associativity condition for the star product is:
\be \a*\(\b*\g\)-\(\a*\b\)*\g=0. \label{assoc}
\ee
To   $\mathcal{O}\(\hbar^2\)$, we have
\align{
\a*\(\b*\g\)&=\a\b\g + \hbar \( \a\{ \b,\g\} +  \{\a, \b\g\}\) + \hbar^2 \( \{\a, \{\b,\g\}\} + \a C_2(\b,\g) +  C_2(\a, \b\g) \) \\
\(\a*\b\)*\g&=\a\b\g + \hbar \( \{\a, \b\}\g + \{\a\b,\g \} \) + \hbar^2 \( \{\{\a,\b\},\g\} + C_2(\a,\b)\g +  C_2(\a\b,\g) \).
}
The associativity condition is trivially satisfied at $\mathcal{O}\(1\)$.  At $\mathcal{O}\(\hbar\)$, we must use the graded product rule of the Poisson bracket: 
$$\{\a\b,\g \}= \a\{\b, \g\} + (-1)^{|\b||\g|} \{\a, \g\}\b.$$
Checking the associativity condition at order $\hbar$, we find:
\align{ \a *(\b *\g)|_{\hbar}-(\a *\b)*\g|_{\hbar} &=a\{\b,\g\}+\{\a,\b\g\} -
  \{\a,\b\}\g - \{\a\b,\g\} 
  \\ &=\a\{\b,\g\}+\{\a,\b\}\g+(-1)^{|\a||\b|}\b\{\a,\g\} -\{\a,\b\}\g
  - \a\{\b,\g\} + (-1)^{|\b||\g|}\{\a,\g\}\b, 
} 
where the last line makes it evident that the associativity condition is satisfied.  

At $\mathcal{O}\(\hbar^2\)$, the calculation rapidly becomes more
complicated.  The form of Equation \ref{assoc} at $\mathcal{O}(\hbar^2)$ gives:
$$ \{\a, \{\b,\g\}\} + \a C_2(\b,\g) + C_2(\a, \b\g) -\{\{\a,\b\},\g\} - C_2(\a,\b)\g - C_2(\a\b,\g)=0.$$
For brevity, we rewrite this condition as:
$$\delta C_2(\a,\b,\g)\equiv \a C_2(\b,\g)-C_2(\a\b,\g)+C_2(\a,\b\g)-C_2(\a,\b)\g = \{\{\a,\b\},\g\} - \{\a,\{\b,\g\}\},$$
where $\delta$ is called the Hochschild coboundary operator. We do not go
into details of Hoschschild cohomology, but this is a useful shorthand
notation.

To check the associativity condition for $C_2(\a,\b)$, we use the graded Jacobi
identity to rewrite $\{\{\a,\b\},\g\}$:
$$\{\{\a,\b\},\g\} = \{\a, \{\b,\g\}\}  +(-1)^{|\a||\b| +1}\{\b, \{\a, \g\}\}.$$
Using the Poisson bracket (Equation \ref{Ansatz}) to calculate $\{\{\a,\b\},\g\}$ and $\{\{\a,\b\},\g\}$,
{\small
\begin{eqnarray*}
\lefteqn{(-1)^{|\a||\b| +1}\{\b,\{\a,\g\}\} = -\pi^{ab}\bigg[\d_b \pi^{mn}\nabla_{e_m}\a \nabla_{e_a}\b \nabla_{e_n}\g +\pi^{mn}\(\nabla_{e_b}\nabla_{e_m}\a \nabla_{e_a}\b \nabla_{e_n}\g +\nabla_{e_m}\a \nabla_{e_a}\b\nabla_{e_b}\nabla_{e_n}\g \) \bigg]} && \nonumber \\
&& \phantom{(-1)^{|\a||\b| +1}\{\b,\{\a,\g\}\} =} +\pi^{ab}\t{R}^{mn} \Bigg[ (-1)^{|\a|+|\b|} (i_{e_m}\nabla_{e_a}\a) (\nabla_{e_b}\b) ( i_{e_n}\g) + (-1)^{|\a|+|\b|-1} (i_{e_m}\a)(\nabla_{e_a}\b)  (i_{e_n} \nabla_{e_b}\g) \nonumber \\
&& \phantom{(-1)^{|\a||\b| +1}\{\b,\{\a,\g\}\} =+\pi^{ab}\t{R}^{mn} \Bigg[}+ (-1)^{|\a|} (i_{e_n}\nabla_{e_a}\a) (i_{e_m}\b)(\nabla_{e_b}\g)+(-1)^{|\b|-1} (\nabla_{e_a} \a) (i_{e_m}\b)(i_{e_n}\nabla_{e_b} \g) \Bigg]  \nonumber \\
&& \phantom{(-1)^{|\a||\b| +1}\{\b,\{\a,\g\}\} =}-\t{R}^{ab}\t{R}^{mn}\bigg[ (-1)^{|\b|} (i_{e_b} i_{e_m}\a)(i_{e_a}\b)( i_{e_n}\g) +(-1)^{|\a|}(i_{e_m}\a)(i_{e_a}\b)  (i_{e_b} i_{e_n}\g)\bigg]  \nonumber \\
&& \phantom{(-1)^{|\a||\b| +1}\{\b,\{\a,\g\}\} =}+(-1)^{|\b|-1}\t{R}^{ab} (i_{e_b}\t{R}^{mn})~ (i_{e_m}\a)(i_{e_a}\b) (i_{e_n}\g).
\end{eqnarray*}
}
The three terms in the first line do not contain $\t{R}^{ij}$, while the remaining terms do contain $\t{R}^{ij}$.  This means that the associativity condition at $\mathcal{O}\(\hbar^2\)$ separates into two pieces:
\be
 \delta C_2^\nabla(\a,\b,\g) \overset{?}{=}-\pi^{ab}\(\d_b \pi^{mn}\nabla_{e_m}\a \nabla_{e_a}\b \nabla_{e_n}\g +\pi^{mn}\(\nabla_{e_b}\nabla_{e_m}\a \nabla_{e_a}\b \nabla_{e_n}\g +\nabla_{e_m}\a \nabla_{e_a}\b\nabla_{e_b}\nabla_{e_n}\g \) \)\label{small}
 \ee
and
\begin{eqnarray}
\lefteqn{ \delta C_2^{\t{R}}(\a,\b,\g)\overset{?}{=}\pi^{ab}\t{R}^{mn} \Bigg[ (-1)^{|\a|+|\b|} (i_{e_m}\nabla_{e_a}\a) (\nabla_{e_b}\b) ( i_{e_n}\g) - (-1)^{|\a|+|\b|} (i_{e_m}\a)(\nabla_{e_a}\b)  (i_{e_n} \nabla_{e_b}\g) }&& \nonumber\\
&& \phantom{\delta C_2^{\t{R}}(\a,\b,\g)=\pi^{ab}\t{R}^{mn} \Bigg[ }+ (-1)^{|\a|} (i_{e_n}\nabla_{e_a}\a) (i_{e_m}\b)(\nabla_{e_b}\g)+(-1)^{|\b|-1} (\nabla_{e_a} \a) (i_{e_m}\b)(i_{e_n}\nabla_{e_b} \g) \Bigg] \nonumber\\
 && \phantom{\delta C_2^{\t{R}}(\a,\b,\g)=}+\t{R}^{ab}\t{R}^{mn}\bigg[ (-1)^{|\b|} (i_{e_b} i_{e_m}\a)(i_{e_a}\b)( i_{e_n}\g) +(-1)^{|\a|-1}(i_{e_m}\a)(i_{e_a}\b)  (i_{e_b} i_{e_n}\g)\bigg] \nonumber\\
&& \phantom{\delta C_2^{\t{R}}(\a,\b,\g)=}+(-1)^{|\b|-1}\t{R}^{ab} (i_{e_b}\t{R}^{mn})~ (i_{e_m}\a)(i_{e_a}\b) (i_{e_n}\g).\label{big}
\end{eqnarray}

\subsection{Associativity of $C_2^\nabla$}

To show that $C_2^\nabla$ satisfies the associativity condition, we calculate the following quantity:
{\small
\begin{eqnarray*}
\lefteqn{C_2^\nabla(\a\b,\g)-\a C_2^\nabla(\b,\g) =\frac12 \pi^{ij}\pi^{mn}\( \(\nabla_{e_i}\nabla_{e_m}\a\)\b +2 \nabla_{e_i}\a\nabla_{e_m}\b \)\nabla_{e_j}\nabla_{e_n}\g } &&\\
&&\phantom{C_2^\nabla(\a\b,\g)-\a C_2^\nabla(\b,\g)}+\frac13\pi^{ia}\d_a\pi^{mn}\bigg[\big(\(\nabla_{e_i}\nabla_{e_m}\a\)\b
+ \nabla_{e_i}\a\nabla_{e_m}\b +\nabla_{e_m}\a\nabla_{e_i}\b \big)\nabla_{e_n}\g-\(\nabla_{e_m}\a\)\b\nabla_{e_i}\nabla_{e_n}\g\bigg].
\end{eqnarray*}
}
This is found via a straightforward calculation. The quantity $C_2^\nabla(\a,\b\g)-C_2^\nabla(\a,\b)\g$ can be calculated similarly, so that $\delta C_2^\nabla$ is:
\begin{eqnarray*}
\lefteqn{\delta C_2^\nabla(\a,\b,\g)= \pi^{ij}\pi^{mn}\bigg[\(\nabla_{e_i}\nabla_{e_m}\a\)\nabla_{e_j}\b\nabla_{e_n}\g - \nabla_{e_i}\a\nabla_{e_m}\b\(\nabla_{e_j}\nabla_{e_n}\g\)  \bigg] }&& \\
&& \phantom{\delta C_2^\nabla(\a,\b)=} -\frac13 \pi^{ij}\d_j \pi^{mn}\bigg[ \nabla_{e_m}\a\(\nabla_{e_i}\b\nabla_{e_n}\g +\nabla_{e_n}\b\nabla_{e_i}\g\) + \( \nabla_{e_i}\a\nabla_{e_m}\b +\nabla_{e_m}\a\nabla_{e_i}\b \)\nabla_{e_n}\g  \bigg].
\end{eqnarray*}

Notice that the second term can be rewritten as:
$$-\frac13 \bigg[ \pi^{mj}\d_j\pi^{in}+\underset{=-\pi^{mj}\d_j\pi^{ni}}{\underbrace{\pi^{nj}\d_j\pi^{im}+\pi^{ij}\d_j\pi^{mn}}}+\pi^{mj}\d_j\pi^{in} \bigg]\nabla_{e_i}\a\nabla_{e_m}\b \nabla_{e_n}\g =- \pi^{mj}\d_j\pi^{in} \nabla_{e_i}\a\nabla_{e_m}\b \nabla_{e_n}\g $$
The terms in the middle combine via the Jacobi identity. Thus we have the following expression for $\delta C_2^\nabla$:
$$\delta C_2^\nabla(\a,\b,\g)= \pi^{ij}\pi^{mn}\bigg[\(\nabla_{e_i}\nabla_{e_m}\a\)\nabla_{e_j}\b\nabla_{e_n}\g - \nabla_{e_i}\a\nabla_{e_m}\b\(\nabla_{e_j}\nabla_{e_n}\g\)  \bigg] -\pi^{mj}\d_j\pi^{ni} \nabla_{e_i}\a\nabla_{e_m}\b\nabla_{e_n}\g.$$
This is identitcal to the right-hand side of (\ref{small}), so $C_2^\nabla$ is associative at order $\hbar^2$.

\subsection{Associativity of $C_2^{\t{R}}$}

Calculating the associativity  $C_2^{\t{R}}$ in the same way, we construct $\delta C_2^{\t{R}}$ by finding:
{\small
\begin{eqnarray*}
\a C_2^{\t{R}}(\b,\g) = -\frac12 \t{R}^{ab}\t{R}^{mn}\a (i_{e_a}i_{e_m}\b)(i_{e_b}i_{e_n} \g) -\frac13 \t{R}^{ab}(i_{e_b}\t{R}^{mn})\bigg[ (-1)^{|\a|+|\b|}\a (i_{e_a}i_{e_m}\b)(i_{e_n}\g)+(-1)^{|\a|}\a (i_{e_m}\b)(i_{e_a}i_{e_n}\g)\bigg] \\
+(-1)^{|\b|}\pi^{ab}\t{R}^{mn}\a(i_{e_m}\nabla_{e_a} \b)(i_{e_n}\nabla_{e_b}\g)
\end{eqnarray*}
}
and
{\small
\begin{eqnarray*}
\lefteqn{C_2^{\t{R}}(\a,\b\g) = -\frac12 \t{R}^{ab}\t{R}^{mn}(i_{e_a}i_{e_m}\a)\bigg[ (i_{e_b}i_{e_n}\b)\g +(-1)^{|\b|-1}(i_{e_n}\b)(i_{e_b}\g)+(-1)^{|\b|} (i_{e_b}\b)(i_{e_n}\g) + \b(i_{e_b}i_{e_n}\g)\bigg]} \\
&& \phantom{C_2^{\t{R}}(\a,\b\g) =} -\frac13 \t{R}^{ab}(i_{e_b}\t{R}^{mn})\bigg[ (-1)^{|\a|} (i_{e_a}i_{e_m}\a)(i_{e_n}\b)\g +(-1)^{|\a|+|\b|}(i_{e_a}i_{e_m}\a)\b(i_{e_n}\g)+(i_{e_m}\a)(i_{e_a}i_{e_n}\b)\g \\
&& \phantom{C_2^{\t{R}}(\a,\b\g) =  -\frac13 \t{R}^{ab}(i_{e_b}\t{R}^{mn})}+(-1)^{|\b|-1}(i_{e_m}\a)(i_{e_n}\b)(i_{e_a}\g)+(-1)^{|\b|}(i_{e_m}\a) (i_{e_a}\b) (i_{e_n}\g) + (i_{e_m}\a)\b(i_{e_a}i_{e_n}\g)\bigg]\\
&& \phantom{C_2^{\t{R}}(\a,\b\g) =}+(-1)^{|\a|}\pi^{ab}\t{R}^{mn}(i_{e_m}\nabla_{e_a}\a)\bigg[ (i_{e_n}\nabla_{e_b}\b)\g +(i_{e_n}\b)\nabla_{e_b}\g +(-1)^{|\b|}\nabla_{e_b}\b (i_{e_n}\g) +(-1)^{|\b|} \b (i_{e_n}\nabla_{e_b}\g)\bigg].
\end{eqnarray*}
}
We get similar equations for $C_2^{\t{R}}(\a\b,\g)$ and
$C_2^{\t{R}}(\a,\b)\g$. Combining everything, we get:
\begin{eqnarray*}
\lefteqn{\delta C_2^{\t{R}}(\a,\b,\g)=-\frac12 \t{R}^{ab}\t{R}^{mn}\bigg[2(-1)^{|\a|}(i_{e_m}\a)(i_{e_n}\b)(i_{e_b}i_{e_n}\g)+2(-1)^{|\b|}(i_{e_a}i_{e_m}\a)(i_{e_b}\b)(i_{e_n}\g)\bigg]} && \\
&&\phantom{\delta C_2^{\t{R}}(\a,\b,\g)=}+\pi^{ab}\t{R}^{mn}\bigg[(-1)^{|\a|+|\b|+1}(i_{e_m}\a)\nabla_{e_a}\b(i_{e_n}\nabla_{e_b}\g)+(-1)^{|\b|+1}\nabla_{e_a}\a (i_{e_m}\b)(i_{e_n}\nabla_{e_b}\g) \\
&& \phantom{\delta C_2^{\t{R}}(\a,\b,\g)=+\pi^{ab}\t{R}^{mn}\bigg[}+(-1)^{|\a|}(i_{e_m}\nabla_{e_a}\a)(i_{e_n}\b)(\nabla_{e_b}\g) +(-1)^{|\a|+|\b|}(i_{e_m}\nabla_{e_a}\a)\nabla_{e_b}\b(i_{e_n}\g)\bigg] \\
&&\phantom{\delta C_2^{\t{R}}(\a,\b,\g)=} -\frac13(-1)^{|\b|}\bigg[ 2\t{R}^{ab}(i_{e_b}\t{R}^{mn})\underset{=+\t{R}^{ab}(i_{e_b}\t{R}^{mn})}{\underbrace{-\t{R}^{mb}(i_{e_b}\t{R}^{an})-\t{R}^{nb}(i_{e_b}\t{R}^{ma})}}\bigg](i_{e_m}\a)(i_{e_a}\b)(i_{e_n}\g).
\end{eqnarray*}
This is identical to the right-hand side of (\ref{big}), so $C_2^{\t{R}}$ is associative at order $\hbar^2$.  Combining the result for $C_2^{\t{R}} $ with the result for $C_2^{\nabla}$, this calculation shows $C_2$ is associative at order $\hbar^2$.  Thus, the proposed product is associative, and therefore a star product.


\begin{thebibliography}{99}

\bibitem{f} 
  F.~Bayen, M.~Flato, C.~Fronsdal, A.~Lichnerowicz and D.~Sternheimer,
  ``Deformation Theory And Quantization. 1. Deformations Of Symplectic Structures,''
  Annals Phys.\  {\bf 111}, 61 (1978).
  \\ F.~Bayen, M.~Flato, C.~Fronsdal, A.~Lichnerowicz and D.~Sternheimer,
  ``Deformation Theory And Quantization. 2. Physical Applications,''
  Annals Phys.\  {\bf 111}, 111 (1978).


\bibitem{cf}
  A.~S.~Cattaneo and G.~Felder,
  ``A path integral approach to the Kontsevich quantization formula,''
  Commun.\ Math.\ Phys.\  {\bf 212}, 591 (2000)
  [arXiv:math/9902090].

\bibitem{ch} 
  C.~S.~Chu and P.~M.~Ho,
  ``Poisson Algebra Of Differential Forms,''
  Int.\ J.\ Mod.\ Phys.\  {\bf 12}, 5573 (1997)
  [arXiv:q-alg/9612031].


\bibitem{ds}
  G.~Dito and D.~Sternheimer,
  ``Deformation Quantization: Genesis, Developments and Metamorphoses,''
  arXiv:math/0201168.


\bibitem{jf}
  J.~Fuchs,
  ``Affine Lie algebras and quantum groups: An Introduction, with applications in conformal field theory,''
  {\it  Cambridge, UK: Univ. Pr. (1992) 433 p. (Cambridge monographs on mathematical physics). 1st publ.}




\bibitem{hm}
  P.~M.~Ho and S.~P.~Miao,
  ``Noncommutative differential calculus for D-brane in non-constant B  field background,''
  Phys.\ Rev.\  D {\bf 64}, 126002 (2001)
  [arXiv:hep-th/0105191].



\bibitem{k}
  M.~Kontsevich,
  ``Deformation quantization of Poisson manifolds, I,''
  Lett.\ Math.\ Phys.\  {\bf 66}, 157 (2003)
  [arXiv:q-alg/9709040].
  



\bibitem{kv}   
  V.~G.~Kupriyanov and D.~V.~Vassilevich,
  ``Star products made (somewhat) easier,''
  arXiv:0806.4615 [hep-th].



\bibitem{sw}
N.~Seiberg and E.~Witten,
  JHEP {\bf 9909}, 032 (1999)
  [arXiv:hep-th/9908142].



\bibitem{v} I. Vaisman, ``Lectures on the Geometry of Poisson Manifolds,'' 
{\it Berlin, Germany: Birkh\"auser Verlag, (1994) 205 p. (Progress in mathematics; Vol. 118). 1st publ.}



\end{thebibliography}
\end{document}